\documentclass[amsmath,amssymb,aps,prd,reprint,twocolumn,showkeys,showpacs]{revtex4-1}

\usepackage{xcolor}
\usepackage{amsmath}
\usepackage{graphicx}
\usepackage{dcolumn}
\usepackage{bm}
\usepackage{epstopdf}
\printfigures
\usepackage{epsfig}
\usepackage[]{quoting}
\usepackage{subfigure}

\usepackage{booktabs}

\begin{document}

\title{Dark effects in $\Tilde{f}(R,P)$ gravity}

\author{Mihai Marciu}
\email{mihai.marciu@drd.unibuc.ro}
\affiliation{Faculty of Physics, University of Bucharest, 405 Atomi\c{s}tilor, POB MG-11, RO-077125, Bucharest-M\u{a}gurele, Romania}

\begin{abstract}
In the present paper a new cosmological model is proposed by extending the Einstein--Hilbert Lagrangian with a generic functional $\Tilde{f}(R,P)$, which depends on the scalar curvature $R$ and a term $P$ which encodes a possible influence from specific cubic contractions of the Riemann tensor. After proposing the corresponding action, the associated modified Friedmann relations are deduced, in the case where the generic functional has the following decomposition, $\Tilde{f}(R,P)=f(R)+g(P)$. The present study takes into account the power--law and the exponential decomposition for the specific form of the corresponding generic functional. For the analytical approach the specific method of dynamical system analysis is employed, revealing the fundamental properties of the phase space structure, discussing the dynamical consequences for the cosmological solutions obtained. It is revealed that the cosmological solutions associated to the critical points can explain various dynamical eras, with a high sensitivity to the values of the corresponding  parameters, encoding different effects due to the geometrical nature of the specific couplings.
\end{abstract}

\maketitle

\newpage 

\section{Introduction}
\par
The accelerated expansion of the Universe represents an important fundamental question in modern physics. The solution to this question can offer new insights for the gravitational theory, leading to a golden era for the cosmological context, opening various directions in modern physics. The phenomenon behind the accelerated expansion of the Universe represents an enigma, having different implications to the development of science and technology. This phenomenon has been studied intensively in the past two decades, being supported by various different observational analyses \cite{Li_2011, Copeland:2006wr, doi:10.1146/annurev.astro.46.060407.145243,PhysRevD.64.123527,PhysRevLett.124.061101,PhysRevD.101.084049,10.1093/mnras/stx2820,Visinelli:2019qqu}.
\par 
The modified gravity theories \cite{CLIFTON20121, Nojiri:2017ncd, Nojiri:2010wj, Nojiri:2006ri, Bamba:2012cp} represent an attempt of correcting the basic $\Lambda CDM$ model, a theoretical approach which aims towards a more complete framework capable of solving different inconsistencies \cite{DOLGOV20031, Koyama_2016, Dutta2020, Wang:2018fng, Nojiri:2003ft, Vagnozzi:2019kvw} of the present cosmological scenarios. These theories are based on specific modifications or replacements of the Einstein--Hilbert action, leading to a dynamical evolution of the dark energy equation of state. The main aim of these theories is to explain the known evolution of the Universe from early times to the present days. Within these theories a particular direction has been established recently which include higher order terms in the corresponding action, a specific model which further replaces or complements the basic gravitational theory based on the Einstein--Hilbert action \cite{Bueno:2016ypa}.
\par 
The higher order gravities \cite{Bueno:2016ypa} represent an alternative approach in the modified gravity theories, capable of explaining various aspects and effects for the gravitational interaction. In this framework the Einsteinian cubic gravity was proposed \cite{Bueno:2016xff}, a specific theory based on specific contractions of the Riemann tensor in the third order. After this direction has emerged, various authors have investigated different properties of the Einsteinian cubic gravity or in similar approaches, analyzing the black--holes solutions \cite{Bueno:2017sui, Dykaar:2017mba, Ghodsi:2017iee, Chernicoff:2016qrc, Bueno:2016lrh, Poshteh:2018wqy, Feng:2017tev, Burger:2019wkq,Emond:2019crr,  Cano:2019ozf, Cisterna:2018tgx, Fierro:2020wps, Khodabakhshi:2020ddv, Konoplya:2020jgt, Adair:2020vso, KordZangeneh:2020qeg, Frassino:2020zuv, Hennigar:2018hza}. Moreover, for this scenario, the wormholes properties have been addressed in \cite{Mehdizadeh:2019qvc, Mustafa:2020qjo}. In general the cubic gravity term is expected to have a considerable influence especially at early times during the inflationary epoch \cite{Arciniega:2018tnn}. The effects due to the cubic term in the inflationary epoch have been studied intensively in the last years \cite{Arciniega:2018tnn, Arciniega:2018fxj, Arciniega:2019oxa, Edelstein:2020lgv, Quiros:2020eim}. In Ref.~\cite{Edelstein:2020nhg} the authors have investigated the inflationary epoch by including a scalar field in a cubic gravity theory, showing the viability of such an epoch in this scenario. The basic aspects for thermodynamics in the case of a generic cubic-quartic gravity have been investigated recently \cite{Mir:2019rik}. The fundamental properties of the anisotropic instabilities in the cubic gravity have been also analyzed \cite{Pookkillath:2020iqq}. In this regard, it has been shown that in the Einsteinian cubic extension various specific pathological instabilities might emerge. Recently, in Ref.~\cite{Jimenez:2020gbw} various pathological aspects have been analyzed in the case of Einsteinian cubic gravity and generalised quasi-topological gravity models.
\par 
The extended cubic gravity based on the third order contractions of the Riemann tensor has been proposed recently \cite{Erices:2019mkd}, a theory which further corrects the Einstein--Hilbert action with a generic functional $f(P)$ which encodes specific effects due to the topological invariant $P$. In this case the topological invariant $P$ is based on some specific contractions of the Riemann tensor \cite{Erices:2019mkd}. The dynamical features for the generalized extended cubic gravity have been addressed in \cite{Marciu:2020ysf}, by taking into account two possible configurations for the cubic term which corrects the Einstein--Hilbert action. An alternative proposal has been analyzed in \cite{Quiros:2020uhr} for the Einsteinian cubic gravity, a specific theory which also includes a cosmological constant. The inclusion of the cubic invariant into possible theories which are embedding scalar fields has been investigated in \cite{Marciu:2020ski}, considering a specific coupling for the quintessence or phantom models. Furthermore, different studies have considered a holographic approach for the Einsteinian cubic gravity \cite{Bueno:2018xqc, Jiang:2019kks}. Recently, the Starobinsky's model of inflation has been corrected by adding a cubic component \cite{Cano:2020oaa}. The authors have investigated the slow--roll regime and discussed the possibility of validation, taking into consideration the scalar and tensor perturbations. 
\par 
In the present paper we shall further generalise the cubic extension of the Einstein--Hilbert action \cite{Erices:2019mkd} by including viable effects from the curvature. Hence, in this approach we shall add to the Einstein--Hilbert action the functional $\Tilde{f}(R,P)$, a generalization which takes into account possible geometrical effects due to the curvature and the inclusion of the cubic invariant. In this case the physical features corresponding to the present model shall be investigated by considering specific methods associated to the dynamical system analysis \cite{BAHAMONDE20181}. This methods have been considered in many cosmological scenarios, representing important analytical tools in the study of physical systems \cite{BAHAMONDE20181}.
\par 
The plan of the present paper is the following. In Sec.~\ref{sec:adoua}  we shall describe the action for the current cosmological model, taking into account a simple decomposition for the generic extension $\Tilde{f}(R,P)$. After writing the action, we shall present the associated modified Friedmann relations, obtained by the variation of the action with respect to the inverse metric. Then, in Sec.~\ref{sec:atreia} we adopt the specific power--law parameterization $\Tilde{f}(R,P)=f_0 R^n+g_0 P^m$ with $f_0, g_0$ and $n, m$ as constant parameters. For this specific model we propose the special form of the auxiliary variables, considering the dynamical system analysis in the case of the power--law parameterization. In Sec.~\ref{sec:atreiab} we analyze the exponential decomposition $\Tilde{f(R, P)}=f_0 e^{n R}+ g_0 e ^{m P}$, with $f_0, g_0, n, m$ constant parameters. After we describe the main features of the phase space for the two models we present a short summary and the final concluding remarks in Sec.~\ref{sec:concluzii}.

\section{The action and the field equations}
\label{sec:adoua} 
\par 
In the present study we shall propose a new cosmological model which is described by the following action:
\begin{equation}
\label{actiune}
S=S_m+\int d^4x \sqrt{- \Tilde{g}} \Bigg( \frac{R}{2}+\Tilde{f}(R,P)\Bigg).
\end{equation}
We note that the Einstein--Hilbert action is further extended by adding a generic functional $\Tilde{f}(R,P)$ which depends on the scalar curvature $R$ and an additional term $P$ which encodes specific cubic contractions of the Riemann tensor \cite{Bueno:2016xff}. In this expression, $\Tilde{g}$ represents the determinant of the metric, while $S_m$ is the action corresponding to the matter sector. Notice that in our action \eqref{actiune} the radiation component is neglected since we are interested in late--time dynamics. The additional term $P$ embedded here has the following decomposition \cite{Erices:2019mkd, Bueno:2016xff}:
\begin{multline}
P=\beta_1 R_{\mu\quad\nu}^{\quad\rho\quad\sigma}R_{ \rho\quad\sigma}^{\quad \gamma\quad\delta}R_{\gamma\quad\delta}^{\quad\mu\quad\nu}+\beta_2 R_{\mu\nu}^{\rho\sigma}R_{\rho\sigma}^{\gamma\delta}R_{\gamma\delta}^{\mu\nu}
\\+\beta_3 R^{\sigma\gamma}R_{\mu\nu\rho\sigma}R_{\quad\quad\gamma}^{\mu\nu\rho}+\beta_4 R R_{\mu\nu\rho\sigma}R^{\mu\nu\rho\sigma}+\beta_5 R_{\mu\nu\rho\sigma}R^{\mu\rho}R^{\nu\sigma}
\\+\beta_6 R_{\mu}^{\nu}R_{\nu}^{\rho}R_{\rho}^{\mu}+\beta_7 R_{\mu\nu}R^{\mu\nu}R+\beta_8 R^3,
\end{multline} encoding specific effects due to the cubic contractions of the Riemann tensor. In this formula the $\beta_i (i=1,8)$ components are constant parameters. In what follows we shall consider the case of a Robertson--Walker metric:
\begin{equation}
\label{metrica}
ds^2=-dt^2+a^2(t) \delta_{ju}dx^j dx^u,
\end{equation}
where $a(t)$ represents the corresponding scale factor, and $t$ the specific cosmic time. In this case the Hubble parameter will be denoted with $H=\frac{1}{a} \frac{da}{dt}$. Note that in our approach we have set the spatial curvature index to zero, a specific value which is compatible with various astrophysical observations. Next, we shall consider the following specific relations between the constant parameters $\beta_i, i=1,8$ \cite{Erices:2019mkd, Bueno:2016xff},

\begin{equation}
\beta_7=\frac{1}{12}\big[3\beta_1-24\beta_2-16\beta_3-48\beta_4-5\beta_5-9\beta_6\big],
\end{equation}
\begin{equation}
\beta_8=\frac{1}{72}\big[-6\beta_1+36\beta_2+22\beta_3+64\beta_4+3\beta_5+9\beta_6\big],
\end{equation}
\begin{equation}
\beta_6=4\beta_2+2\beta_3+8\beta_4+\beta_5,
\end{equation}
\begin{equation}
\bar{\beta}=(-\beta_1+4\beta_2+2\beta_3+8\beta_4).
\end{equation}
In this specific case the cubic term is equal to the following expression \cite{Erices:2019mkd}, 
\begin{equation}
\label{PP}
P=6\bar{\beta}H^4 (2H^2+3\dot{H}),
\end{equation}
describing a topological invariant in the four dimensional space--time.
\par
Furthermore, we shall implement the following decomposition for the specific extension of the Einstein--Hilbert action, by considering
\begin{equation}
\label{descompunere}
    \Tilde{f}(R,P)=f(R)+g(P).
\end{equation}
In the previous formula we shall encode the effects due to the curvature couplings in the $f(R)$ part \cite{Odintsov:2017icc}, while the cubic couplings are embedded into the behavior of the $g(P)$ functional. The scalar curvature in the case of the Robertson--Walker metric \eqref{metrica} is equal to 
\begin{equation}
\label{eqrrr}
    R=6(2 H^2+\dot{H}).
\end{equation}
For the matter component, which is assimilated into the dark matter sector, the energy--momentum tensor is defined as:
\begin{equation}
    T_{\mu\nu}^{(m)}=-\frac{2}{\sqrt{-g}}\frac{\delta S_m}{\delta g^{\mu\nu}}.
\end{equation}
If we take into account the Robertson--Walker metric \eqref{metrica}, we have the following representation for the matter energy--momentum tensor: 
\begin{equation}
    T_{\nu}^{\mu (m)}=diag[-\rho_m, +p_m, +p_m, +p_m],
\end{equation}
with $\rho_m$ the density and $p_m$ the pressure, connected through a barotropic equation of state
\begin{equation}
    w_m=\frac{p_m}{\rho_m},
\end{equation}
where $w_m$ is a constant parameter which describes the properties of the dark matter component. For simplicity we shall consider the dust case where  $w_m=0$. This implies that the the matter component can be regarded as a pressure--less gas which is not too dense.   
\par 
If we consider the variation of the action \eqref{actiune} with respect to the inverse metric $g^{\mu\nu}$, for the previous mentioned decomposition \eqref{descompunere}, we obtain the following modified Friedmann relations:
\begin{equation}
    \label{eqfrcstr}
    3 H^2=\rho_m+\rho_f+\rho_g,
\end{equation}

\begin{equation}
\label{eqacc}
    -2 \dot{H}-3 H^2=p_m+p_f+p_g.
\end{equation}
The densities $\rho_f, \rho_g$ and pressures $p_f, p_g$ have the following expressions \cite{Erices:2019mkd}:
\begin{equation}
    \rho_f=-f(R)-6 H^2 \frac{df(R)}{dR}+R \frac{df(R)}{dR}-6 H \dot{R} \frac{d^2 f(R)}{dR^2},
\end{equation}

\begin{equation}
    \rho_g=-g(P)-18 \beta H^4 \Big(H \frac{\partial}{\partial t} - H^2-\dot{H}\Big) \frac{dg(P)}{dP},
\end{equation}

\onecolumngrid
\begin{equation}
    p_f=4 \frac{df(R)}{dR} \dot{H}+2  \partial_t^2 \Bigg(  \frac{df(R)}{dR}\Bigg) -2 H \partial_t \Bigg(  \frac{df(R)}{dR}\Bigg)- \frac{df(R)}{dR} R+f(R)+6 H \partial_t \Bigg(  \frac{df(R)}{dR}\Bigg)+6 \frac{df(R)}{dR} H^2,
\end{equation}

\begin{equation}
    p_g=g(P)+6 \beta H^3 \Big(H \partial_t^2+2(H^2+2 \dot{H})\partial_t-3 H^3-5 H \dot{H}   \Big) \frac{dg(P)}{dP} .
\end{equation}
\twocolumngrid
\par 
For the present cosmological model the dark energy is encoded into the curvature and cubic couplings, having a geometrical nature. From the definitions of the energy densities and pressures, we note that the current scenario satisfies the continuity equation for the matter and dark energy sector. In the above equations $\partial_t$ and the dot $\dot{}$ defines the partial derivative with respect to the cosmic time, while $\partial_t^2$ represents the double time derivative.
\par 
In the case where $f(R)=0$ the present action \eqref{actiune} describes the so--called cubic extension of gravity, proposed in Ref.~\cite{Erices:2019mkd} and studied in the recent years. Furthermore, by considering $g(P)=0$, we obtain a particular extension of the Einstein--Hilbert action which encodes specific effects due to the curvature, the modified $f(R)$ theory of gravity \cite{Sotiriou:2008rp, Odintsov:2017icc}. This framework has been studied into the past years and represents a viable extension, one of the most studied directions in the scalar tensor theories of gravity \cite{Sotiriou:2008rp, Odintsov:2019ofr, Odintsov:2018uaw, Odintsov:2017tbc}.

\section{The power law decomposition}
\label{sec:atreia} 
\par 
In this section we shall analyze the physical features of the present cosmological model by adopting the dynamical system analysis, an important method specific to physical systems. We shall consider that the $\Tilde{f}(R,P)$  functional is decomposed into a direct sum based on a power--law behavior, i.e. $\Tilde{f}(R,P)=f_0 R^n+g_0 P^m$. For this specific model we shall assume that $f_0, g_0$ and $n, m$ are constant parameters. In what follows we shall introduce the following auxiliary dimension--less variables:

\begin{equation}
    s=\frac{\rho_m}{3 H^2 (1+2 \frac{df(R)}{dR})},
\end{equation}

\begin{equation}
    x_1=\frac{f(R)}{3 H^2 (1+2 \frac{df(R)}{dR})},
\end{equation}

\begin{equation}
    x_2=\frac{R \frac{df(R)}{dR}}{3 H^2 (1+2 \frac{df(R)}{dR})},
\end{equation}

\begin{equation}
    x_3=\frac{2 \dot{R} \frac{d^2 f(R)}{d R^2}}{H (1+2 \frac{df(R)}{dR})},
\end{equation}

\begin{equation}
    z=\frac{R}{H^2},
\end{equation}

\begin{equation}
    y_1=\frac{g(P)}{3 H^2 (1+2 \frac{df(R)}{dR})},
\end{equation}

\begin{equation}
    y_2=\frac{6 \beta H^3 \frac{\partial}{\partial t} (\frac{dg(P)}{dP})}{1+2 \frac{df(R)}{dR}}.
\end{equation}

\par 
At this point we note that the choice of the dimension--less variables is not unique. This specific choice is motivated by the form of the first modified Friedmann relation, the constraint equation. If we take into account that $f(R)=f_0 R^n$ with $f_0, n$ constant parameters, then we obtain an additional relation between $x_2$ and $x_1$ variable, $x_2=n x_1$. In this case the dynamical system becomes a six dimensional system with the associated variables $[s, x_1, x_3, z, y_1, y_2]$. Here the first auxiliary variable $s$ is associated to the matter component, acting as an effective density parameter influenced only by the Hubble parameter and the first variation of the scalar curvature coupling, embedded into the $f(R)$ function. Taking into account the existence conditions specific to $f(R)$ gravity theories, we have: $s \geq 0$. The second auxiliary variable which is independent, the $x_1$ term, is associated with the specific form of the scalar curvature coupling, embedded into the $f(R)$ function. Next, the $z$ variable is connected to the specific value of the scalar curvature $R$, balanced by the square of the Hubble parameter. Finally, the $y_1$ variable is associated to the effects due to the cubic component, while in $y_2$ we notice the influence from the variation of the $P$ invariant term.
\par 
We can define the effective equation of state for the present cosmological system,
\begin{equation}
    w_{eff}=\frac{p_m+p_f+p_g}{\rho_m+\rho_f+\rho_g}=-1-\frac{2}{3}\frac{\dot{H}}{H^2}.
\end{equation}
The effective or total equation of state is connected to the value of the $z$ variable since
\begin{equation}
    \frac{\dot{H}}{H^2}=\frac{z-12}{6}.
\end{equation}
This relation appears due to the value of the scalar curvature, valid for the present metric. Taking into account the previous definitions for the auxiliary variables we can rewrite the Friedmann constraint equations as the following expression, reducing the dimensionality of the dynamical system by determining the $s$ variable, 
\begin{equation}
    s=m y_1 \left(-\frac{2}{z-8}-1\right)-(n-1) x_1+x_3+y_1+y_2+1.
\end{equation}

The next step for the dynamical system analysis method assumes the transformation from the cosmic time $t$ to $N$, where $N=log(a)$. In this case we shall obtain an autonomous system of ordinary differential equations described in the following relations:

\onecolumngrid
\begin{equation}
\label{eqa1}
    \frac{dx_1}{dN}=\frac{x_3 z}{6 (n-1)}-\frac{x_1 z}{3}-x_3 x_1+4 x_1, 
\end{equation}

\begin{equation}
    \frac{dx_3}{dN}=\frac{6 (n-1) n x_1 \ddot{R}}{z^2 H^4}+\frac{1}{6} x_3^2 \left(\frac{(n-2) z}{(n-1) n x_1}-6\right)+x_3 \left(2-\frac{z}{6}\right),
\end{equation}

\begin{equation}
    \frac{dz}{dN}=\frac{x_3 z^2}{6 (n-1) n x_1}-\frac{z^2}{3}+4 z, 
\end{equation}

\begin{equation}
    \frac{dy_1}{dN}=\frac{y_2 z}{6 (m-1)}-\frac{4 y_2}{3 (m-1)}-x_3 y_1-\frac{y_1 z}{3}+4 y_1, 
\end{equation}

\begin{equation}
\label{eqa2}
    \frac{dy_2}{dN}=\frac{1}{6} \left(\frac{12 (m-1) m y_1 \ddot{P}}{\beta  (z-8)^2 H^8}+\frac{(m-2) y_2^2 (z-8)}{(m-1) m y_1}+3 y_2 \left(-2 x_3+z-12\right)\right). 
\end{equation}

\twocolumngrid
\par In the dynamical system determined by the equations \eqref{eqa1}--\eqref{eqa2} we notice that we have two components which have to be determined, $\frac{\ddot{R}}{H^4}$ and $\frac{\ddot{P}}{H^8}$. These components are determined by considering the acceleration equation \eqref{eqacc} with the specific expressions for the scalar curvature and the $P$ invariant term, expressed into the relations \eqref{eqrrr} and \eqref{PP}. Hence, the acceleration equation can be rewritten in terms of the auxiliary variables in the following way: 

\onecolumngrid
\begin{multline}
    H^2-\frac{H^2 z}{3}=-\frac{2 m^2 y_1 z \ddot{P}}{\beta  H^6 (z-8)^2 \left(6 n x_1-z\right)}+\frac{2 m y_1 z \ddot{P}}{\beta  H^6 (z-8)^2 \left(6 n x_1-z\right)}+\frac{H^2 y_2^2 z^2}{3 (m-1) m y_1 \left(6 n x_1-z\right)}+\frac{5 H^2 m y_1 z^2}{(z-8) \left(6 n x_1-z\right)}
    \\-\frac{H^2 y_2^2 z^2}{6 (m-1) y_1 \left(6 n x_1-z\right)}+\frac{4 H^2 y_2^2 z}{3 (m-1) y_1 \left(6 n x_1-z\right)}-\frac{42 H^2 m y_1 z}{(z-8) \left(6 n x_1-z\right)}-\frac{8 H^2 y_2^2 z}{3 (m-1) m y_1 \left(6 n x_1-z\right)}-\frac{6 n^2 x_1\ddot{R}}{H^2 z \left(6 n x_1-z\right)}
    \\+\frac{6 n x_1 \ddot{R}}{H^2 z \left(6 n x_1-z\right)}-\frac{2 H^2 y_2 z^2}{3 \left(6 n x_1-z\right)}-\frac{3 H^2 y_1 z^2}{(z-8) \left(6 n x_1-z\right)}+\frac{24 H^2 y_1 z}{(z-8) \left(6 n x_1-z\right)}+\frac{6 H^2 y_2 z}{6 n x_1-z}+\frac{H^2 x_3^2 z^2}{3 (n-1) n x_1 \left(6 n x_1-z\right)}
    \\-\frac{H^2 x_3^2 z^2}{6 (n-1) x_1 \left(6 n x_1-z\right)}+\frac{6 H^2 n x_1}{6 n x_1-z}+\frac{H^2 n x_1 z}{6 n x_1-z}-\frac{3 H^2 x_1 z}{6 n x_1-z}-\frac{2 H^2 x_3 z}{6 n x_1-z}.
\end{multline}
  \twocolumngrid
  \par 
  Another relation between $\ddot{R}$ and $\ddot{P}$ is determined by considering the specific expressions for these invariants in the case of the Robertson--Walker metric. Taking into account the previous mentioned considerations, we obtain:
  \onecolumngrid
  \begin{multline}
      \ddot{P}=1728 \beta  H^8-\frac{640 \beta  H^8 y_2}{m^2 y_1-m y_1}+\frac{\beta  H^8 y_2 z^3}{m^2 y_1-m y_1}-\frac{26 \beta  H^8 y_2 z^2}{m^2 y_1-m y_1}+\frac{224 \beta  H^8 y_2 z}{m^2 y_1-m y_1}+\frac{2 \beta  H^8 x_3 z^2}{n x_1-n^2 x_1}-3 \beta  H^8 z^3+84 \beta  H^8 z^2
      \\-720 \beta  H^8 z+3 \beta  H^4 \ddot{R}.
  \end{multline}
  \twocolumngrid

\begin{figure}[tb]
\includegraphics[width=8cm]{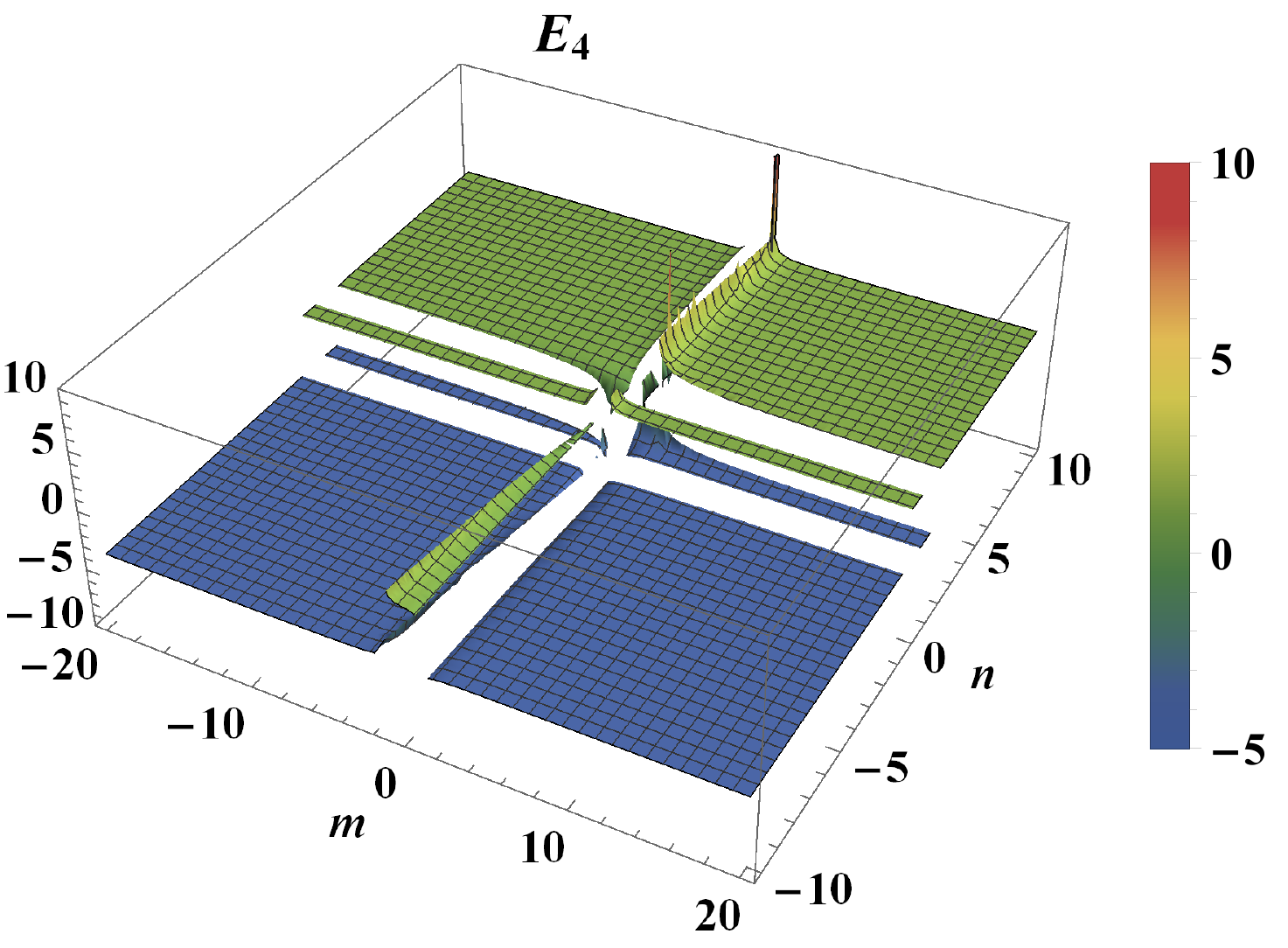}
\centering
\caption{The fourth eigenvalue for the first cosmological solution $A$ in the case where $x_1=1$.}
\label{figE4}
\end{figure}

\begin{figure}[tb]
\includegraphics[width=8cm]{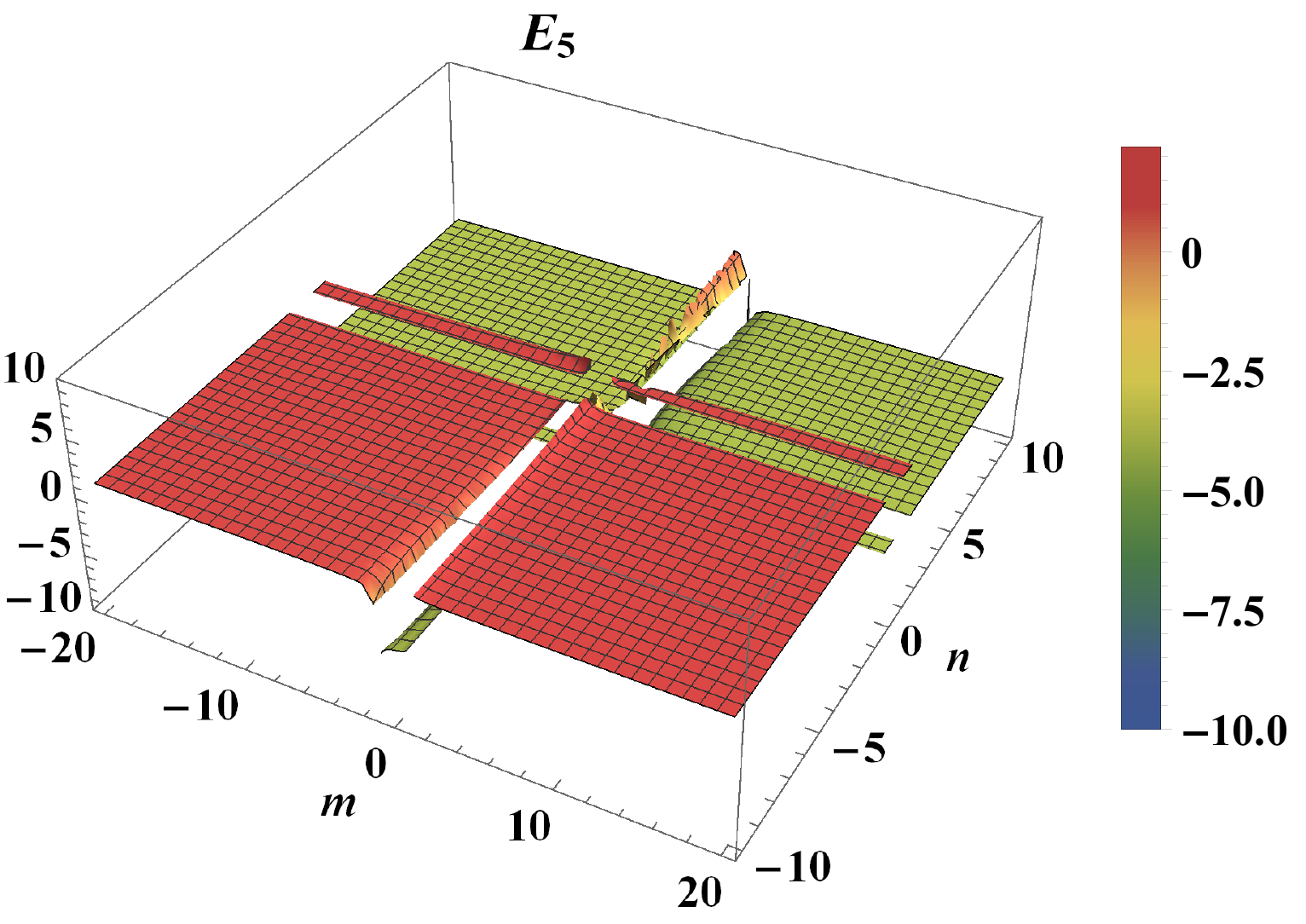}
\centering
\caption{The fifth eigenvalue for the first cosmological solution $A$ in the case where $x_1=1$.}
\label{figE5}
\end{figure}

\par 
At this step we can note that the dynamical system of differential equations \eqref{eqa1}--\eqref{eqa2} is completely autonomous and closed, ready for the dynamical system analysis. For the present model we have obtained two critical points, determined by setting the right hand side of the autonomous equations \eqref{eqa1}--\eqref{eqa2} to zero. 
\par 
The first cosmological solution represents a critical line located at the following coordinates:
\begin{multline}
    A=\Big[x_1, x_3=0, z=12,
    \\y_1=-\frac{2 \left(n x_1-x_1-1\right)}{3 m-2}, y_2=0  \Big].
\end{multline}
For this critical line we note that the $n,m$ parameters are influencing the location in the phase space structure. This critical line is characterized by an indefinite value of the $x_1$ variable which encodes effects due to the scalar curvature coupling. Moreover, the value of the $z$ variable is constant, without any influence from the constant parameters. From a physical point of view this solution describes a de--Sitter epoch where the effective equation of state mimics a cosmological constant, 
\begin{equation}
    w_{eff}=-1.
\end{equation}
In this case the value of the $s$ variable is zero, describing a possible late time stage of the Universe. Hence, this solution can be regarded as a geometrical de--Sitter epoch, with the physical effects encoded into the specific form of the $f(R)$ and $g(P)$ functions which are correcting the Einstein--Hilbert action. From a dynamical perspective this solution is always saddle, due to the specific form of the obtained eigenvalues:

\begin{equation}
    \Big[0, 4, -3, \Big(C_1 \Big)^{-1} \Big(\sqrt{C_2}+C_3\Big), \Big(C_1 \Big)^{-1} \Big(-\sqrt{C_2}+C_3\Big)
    \Big],
\end{equation}
with the following expressions:
\onecolumngrid
\begin{equation}
    C_1=2 (m-1) (3 m-2) (n-1) n x_1 \left((n-1) x_1 \left(18 m^2-3 m (n+6)+2 n\right)-18 (m-1) m\right),
\end{equation}
\begin{multline}
    C_2=(2-3 m)^2 (m-1)^2 (n-1)^2 n^2 x_1^2 ((n-1) x_1 (25 (2-3 m) n+6 m (75 m-59))
    \\-450 (m-1) m-64) ((n-1) x_1 (18 m^2-3 m (n+6)+2 n)-18 (m-1) m),
\end{multline}
\begin{equation}
    C_3=3 (m-1) (3 m-2) (n-1) n x_1 \left((n-1) x_1 (3 m (-6 m+n+6)-2 n)+18 (m-1) m\right).
\end{equation}
\twocolumngrid
For this solution we have displayed in Figs.~\ref{figE4}--\ref{figE5} the specific real values of the fourth and fifth eigenvalues, by setting the value of the $x_1$ variable. At this point, we note that the $n$ and $m$ parameters are only influencing the spiral behavior in the phase space. In this case a spiral trajectory is obtained if the corresponding eigenvalues contain imaginary values. For the $A$ solution a spiral behavior is obtained if the $C_2$ component represents a complex number.

\par
\begin{figure}[t]
\includegraphics[width=8cm]{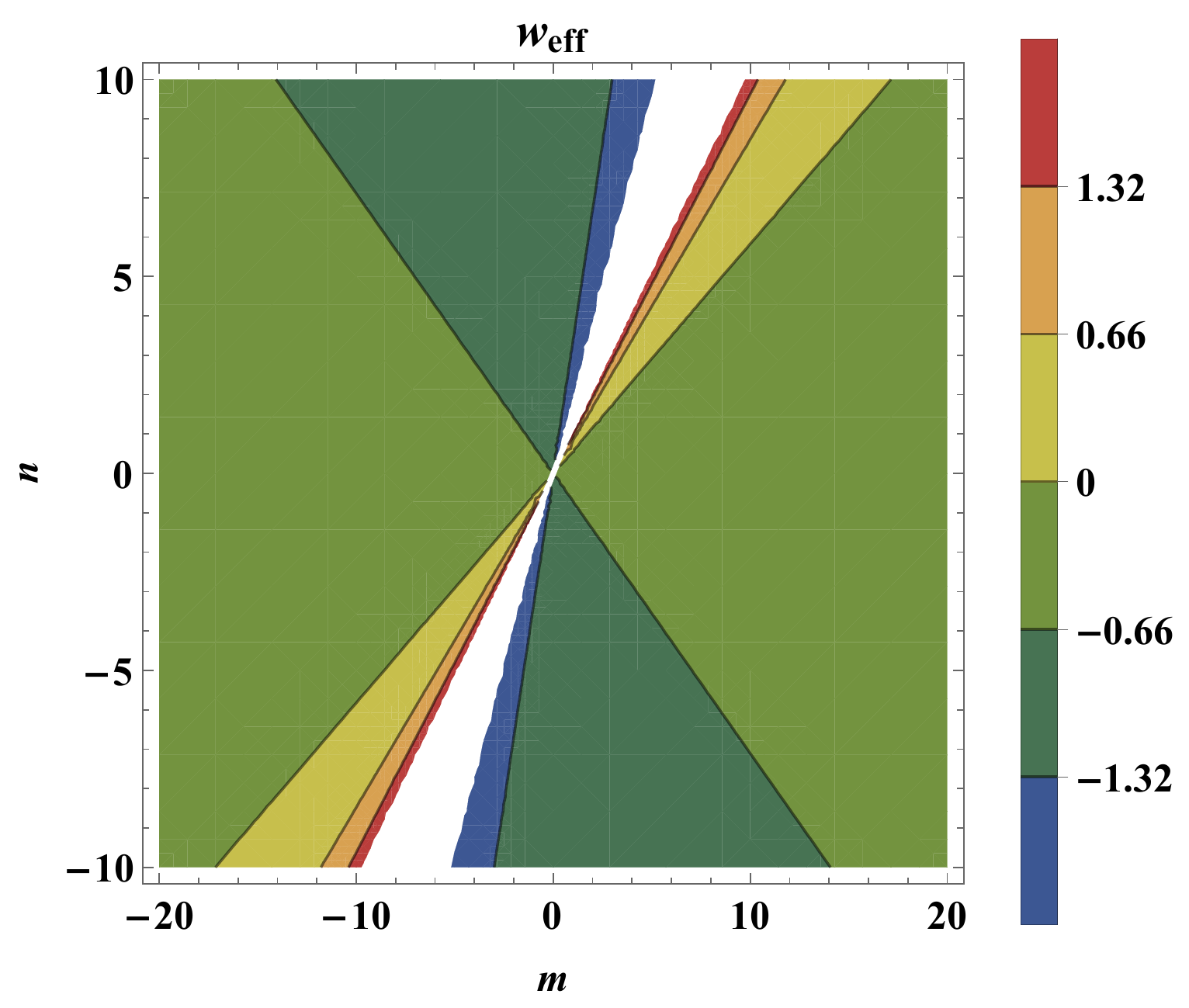}
\centering
\caption{The figure displays the value of the effective equation of state as a function of the two parameters $n$ and $m$ for the $B$ critical point. }
\label{figBweff}
\end{figure}

\begin{figure}[t]
\includegraphics[width=8cm]{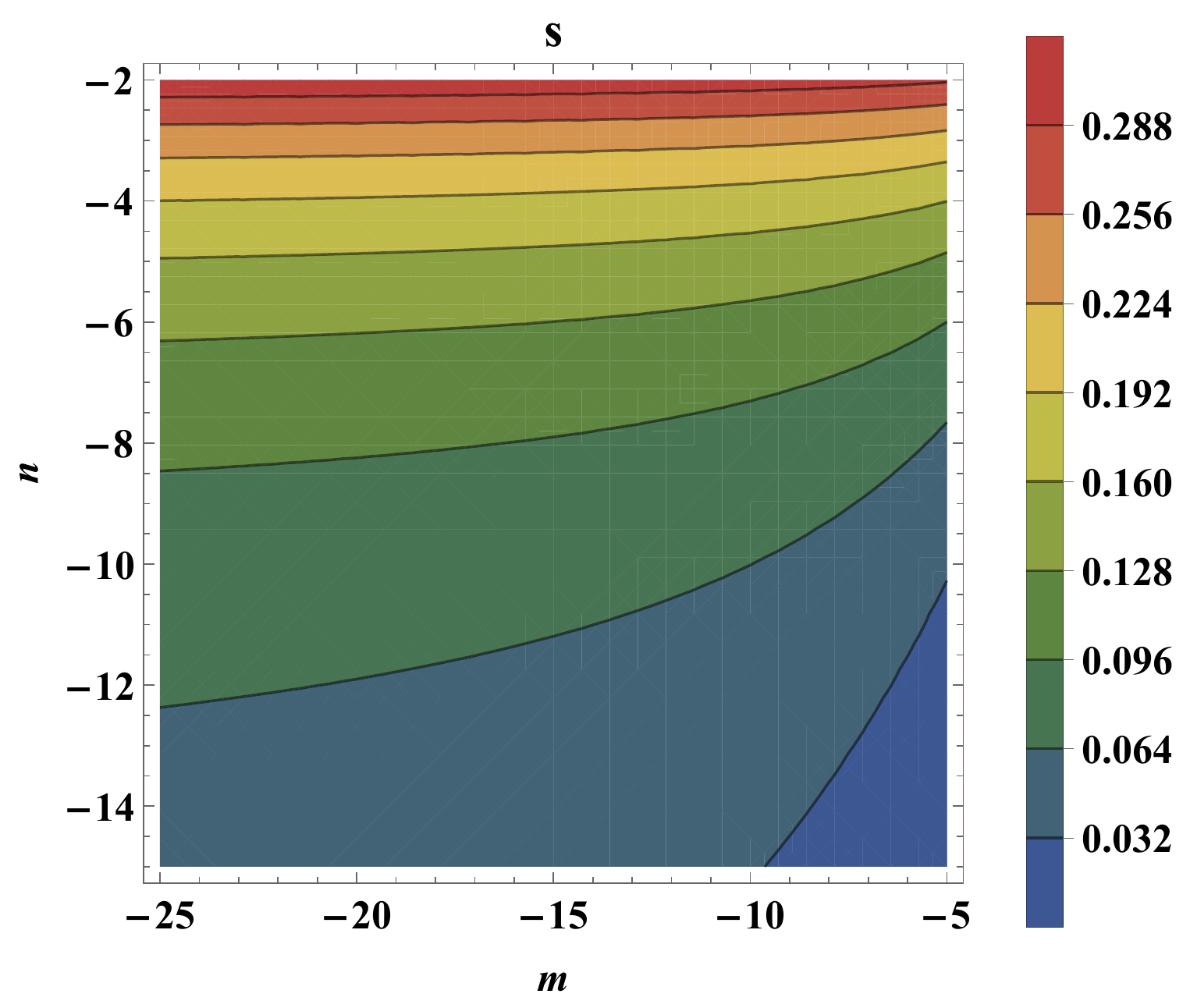}
\centering
\caption{The value of the $s$ variable for the $B$ critical point. For this figure we have displayed only the $[0,1]$ interval.}
\label{figBss}
\end{figure}

\begin{figure}[t]
\includegraphics[width=8cm]{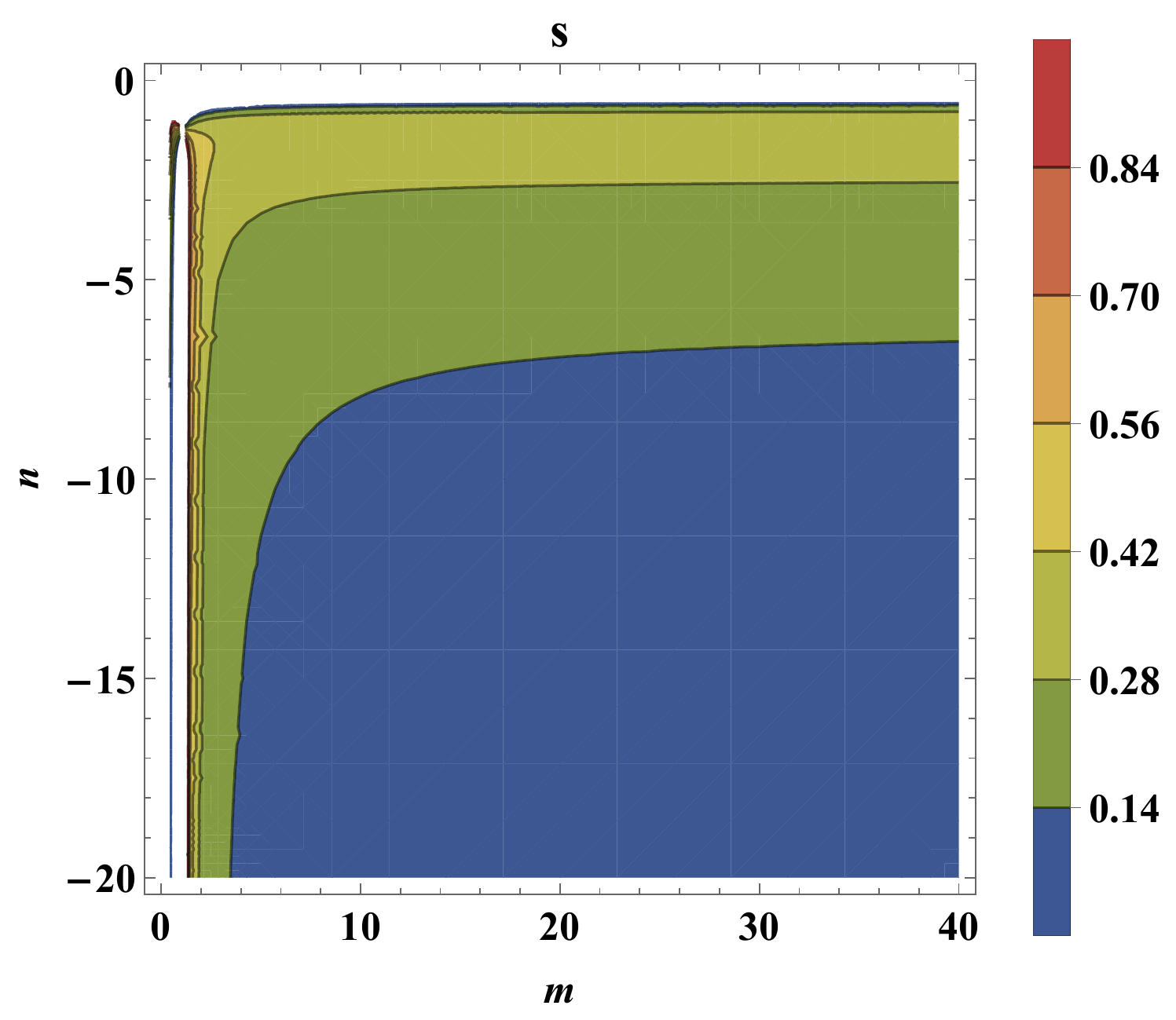}
\centering
\caption{The value of the $s$ variable associated to the $B$ critical point. }
\label{figBsssnew}
\end{figure}

 
The second cosmological solution is located at the following coordinates:
\onecolumngrid
\begin{multline}
    B=\Big[ x_1=\frac{2 (3 m-4 n)}{n (5 m-4 n)}, x_3=-\frac{8 (m n-m)}{5 m-4 n}, z=\frac{12 (3 m-4 n)}{5 m-4 n},
    \\y_1=\frac{2 (m+4 n) (m (8 n-15)+12 n) (m (n (8 n-7)-6)-4 (n-2) n)}{3 n (5 m-4 n) \left(16 \left(4 m^3-7 m+2\right) n^2+5 (5 m-2) m^2-8 (m (14 m-19)+4) m n\right)},
    \\y_2=\frac{8 (m-1) m (m (8 n-15)+12 n) (m (n (8 n-7)-6)-4 (n-2) n)}{(5 m-4 n) \left(16 \left(4 m^3-7 m+2\right) n^2+5 (5 m-2) m^2-8 (m (14 m-19)+4) m n\right) } \Big].
\end{multline}
\twocolumngrid
The effective equations of state presents a sensitivity to the values of the $n$ and $m$ parameters, 
\begin{equation}
    w_{eff}=\frac{12 n-7 m}{3 (5 m-4 n)}.
\end{equation}
In Fig.~\ref{figBweff} we have displayed the graphical representation of the value of the effective equation of state as a function of the $n$ and $m$ parameters. These parameters are encoding the effects due to the curvature and the cubic invariant term, respectively. At this critical point the value of the $s$ variable is the following:
\onecolumngrid
\begin{equation}
s=\frac{8 m \left(-3 m^2 \left(8 n^2-7 n-6\right)+m n \left(8 n^2+5 n-30\right)-4 (n-2) n^2\right)}{3 n \left(m^3 \left(64 n^2-112 n+25\right)+2 m^2 (76 n-5)-16 m n (7 n+2)+32 n^2\right)}.
\end{equation}
\twocolumngrid
The values of the $s$ variable for the second cosmological solution $B$ is displayed in Figs.~\ref{figBss}--\ref{figBsssnew}. We note that in this figure we have displayed only the $[0,1]$ interval. As previously stated, the $s$ component is associated to an effective matter density variable which has to be positive from a physical point of view. Hence, we have displayed in Figs.~\ref{figBss}--\ref{figBsssnew} some possible non--exclusive intervals where the $B$ solution is physically viable. 
\par
\begin{figure}[t]
\includegraphics[width=8cm]{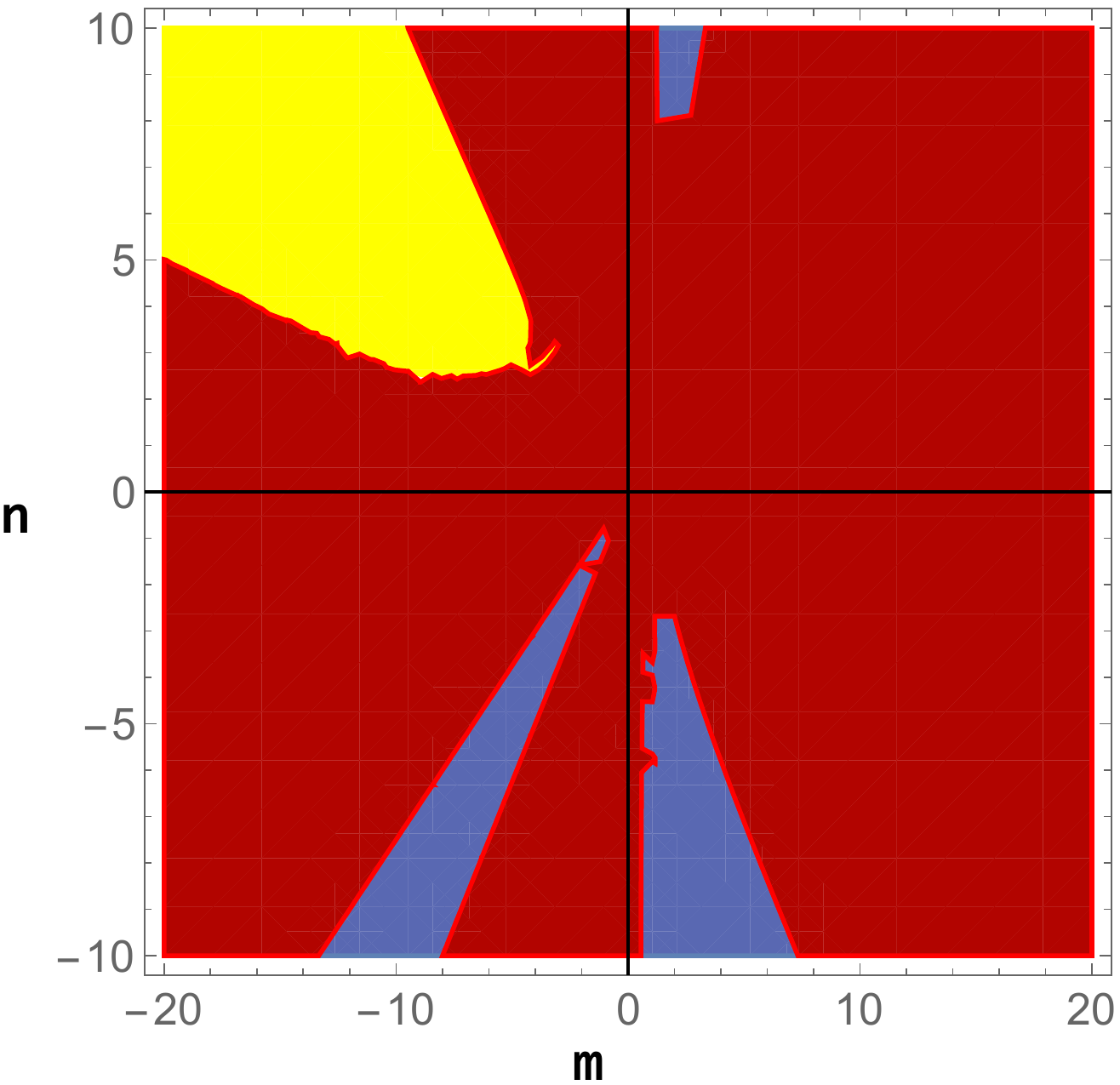}
\centering
\caption{The figure displays the regions in the space associated to $n$ and $m$ parameters for which the $B$ critical point has the following dynamics: unstable (yellow), stable (blue), and saddle (red). }
\label{figBinstabil}
\end{figure}

\begin{figure}[t]
\includegraphics[width=8cm]{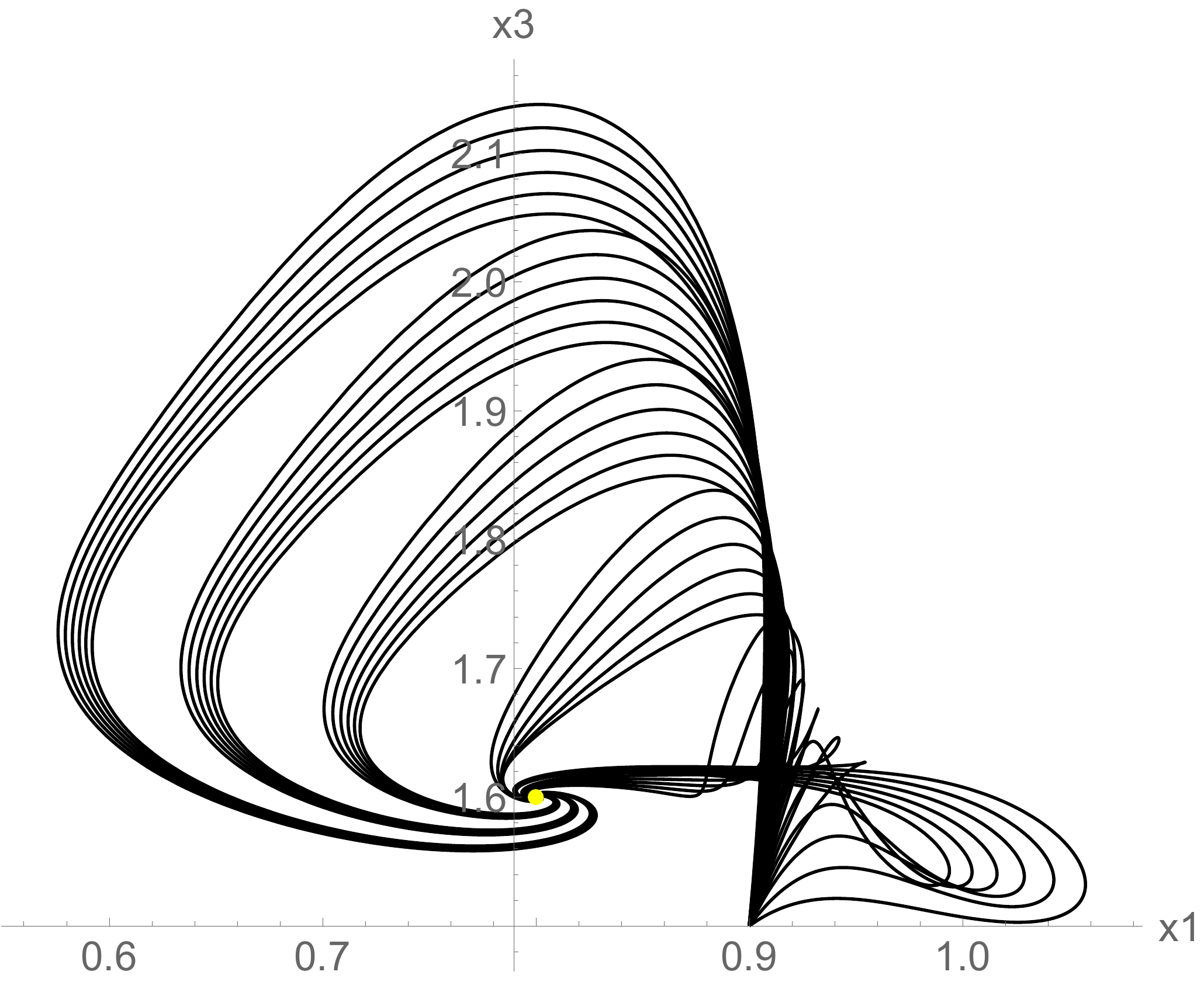}
\centering
\caption{The evolution towards the $B$ critical point in the $x_1 O x_3$ plane in the case where $n=0.5$, $m=0.8$. The initial conditions have been fine--tuned. The $B$ critical point appears as a yellow dot.}
\label{figB1a}
\end{figure}

\begin{figure}[tb]
\includegraphics[width=8cm]{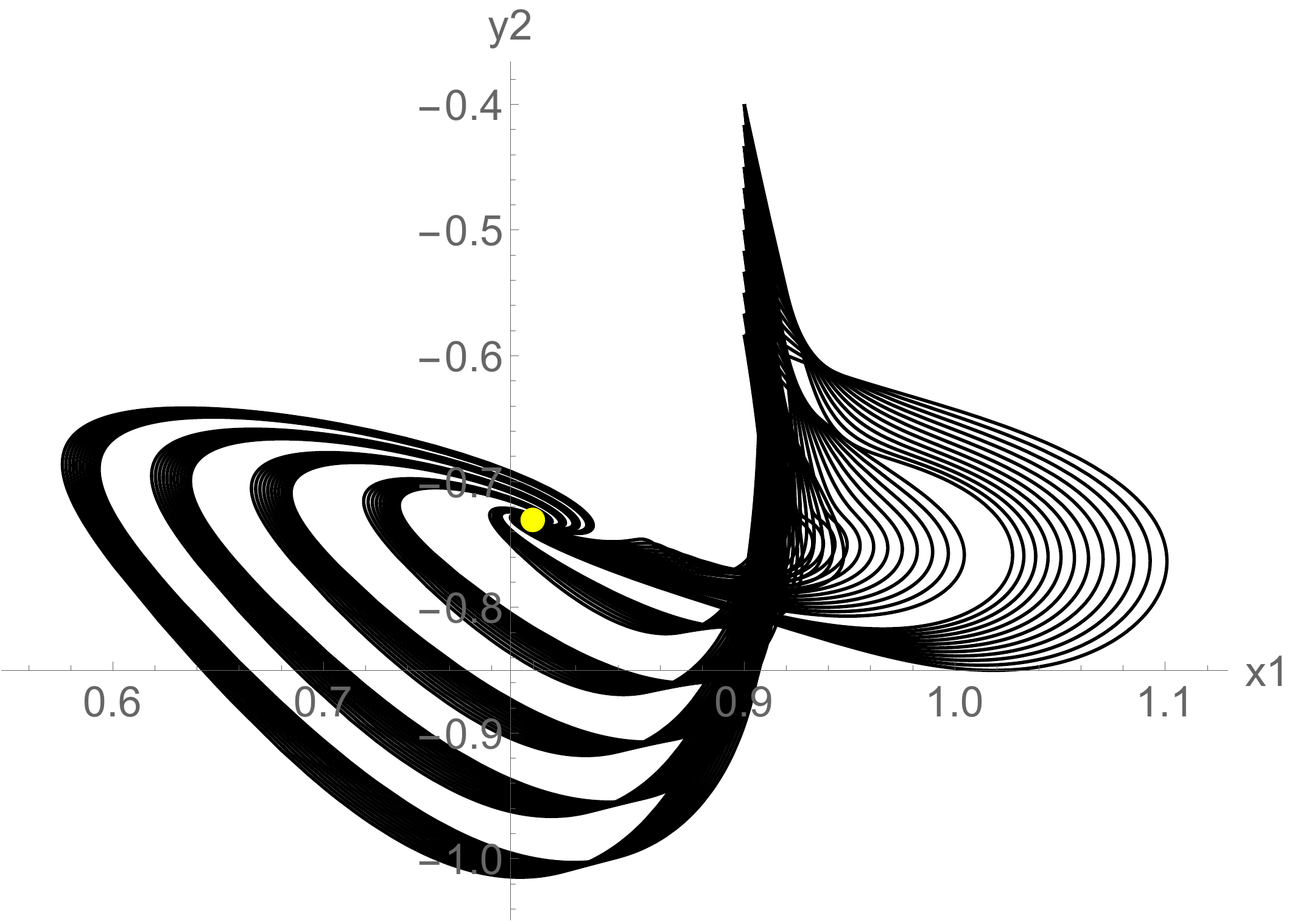}
\centering
\caption{The evolution towards the $B$ critical point in the $x_1 O y_2$ plane for various initial conditions.}
\label{figB1b}
\end{figure}

 
Furthermore, for the dynamical analysis we have obtained the following eigenvalues:
\begin{equation}
    \Big[ \frac{8 m (n-1)}{5 m-4 n}, E_2, E_3, E_4, E_5\Big].
\end{equation}
Due to the complexity, at this point we have written only the expression for the first eigenvalue. The remaining eigenvalues have different complicated expressions and are not written in the manuscript. From the dynamical analysis we have observed that this point can be either stable, saddle or unstable. Hence, we have identified different regions associated to the previous mentioned features, displayed in Fig.~\ref{figBinstabil}. Lastly, for the $B$ critical point, we have shown the viability of the analytical expressions obtained, by considering the numerical evolution in the phase space structure. The numerical evolution towards the $B$ solution is displayed in Figs.~\ref{figB1a}--\ref{figB1b}. We note that from a physical point of view this solution can be associated to various cosmological features, explaining the accelerated expansion effect, as well as other different cosmological epochs in the evolution of the Universe. All of these can be achieved by fine--tuning the values of the $n$ and $m$ parameters.

\begin{figure}[t]
\includegraphics[width=8cm]{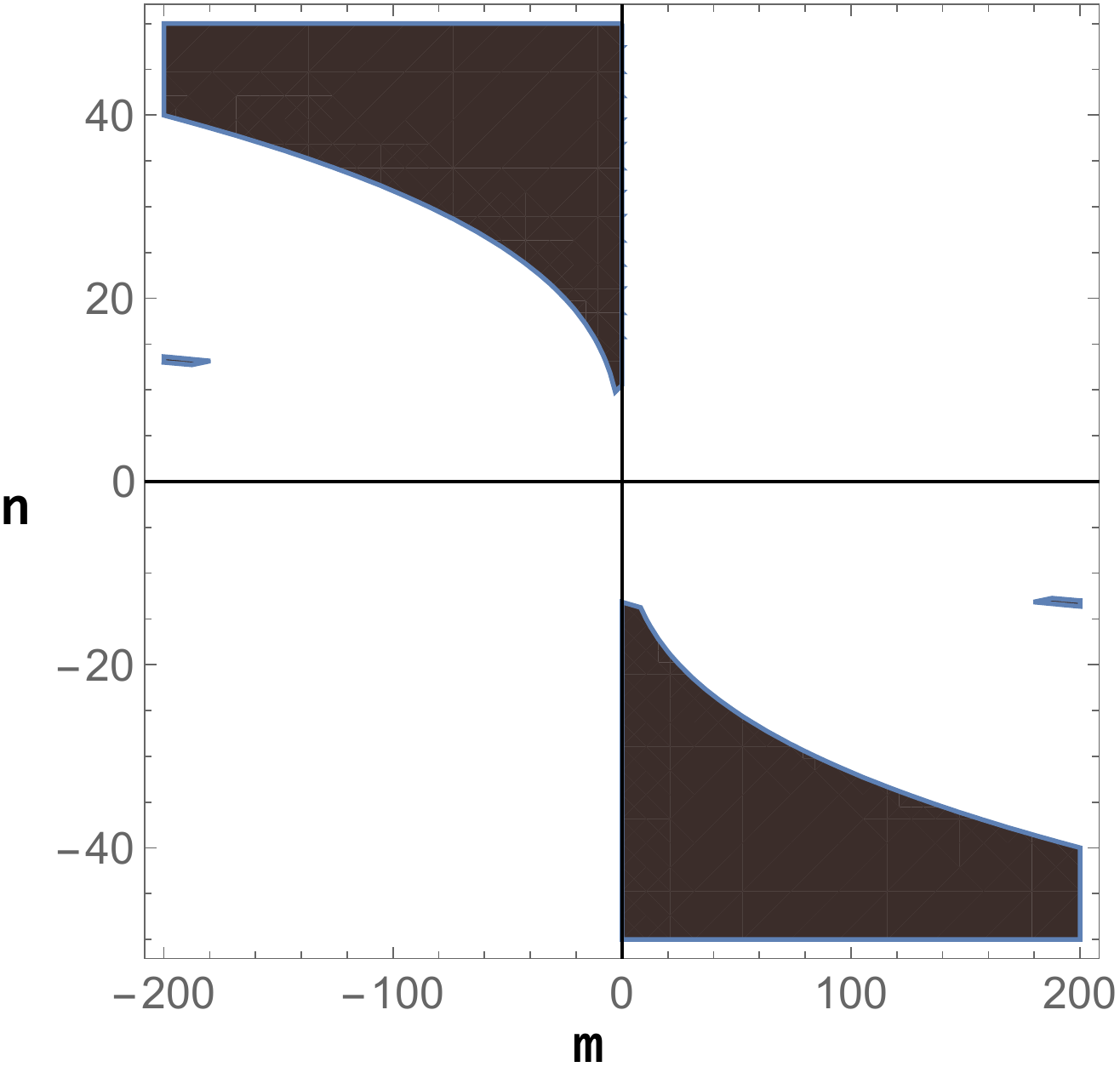}
\centering
\caption{A possible region where the $D$ solution appears to have a spiral behavior due to the imaginary values of the corresponding eigenvalues  $(\beta=-2, x_4=1, y_1=-1)$.}
\label{figDa}
\end{figure}

\begin{figure}[t]
\includegraphics[width=8cm]{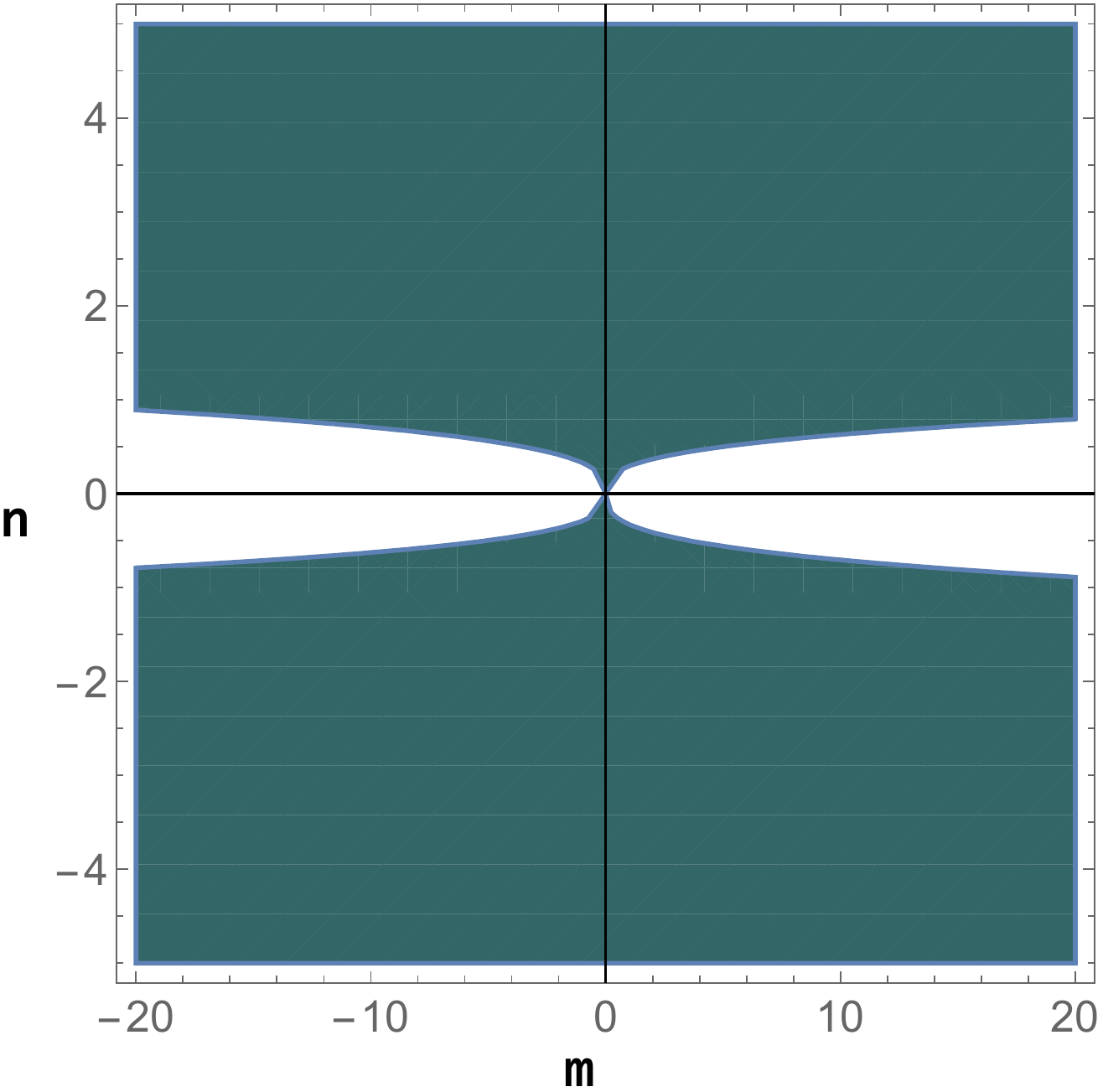}
\centering
\caption{A non--exclusive possible region where the $F$ critical have a spiral behavior due to the imaginary values of the corresponding eigenvalues  $(\beta=-1, x_1=1/2)$.}
\label{figFa}
\end{figure}

\section{The exponential decomposition}
\label{sec:atreiab} 
\par 
In this section we shall adopt a second parameterization which involves a specific exponential decomposition where $\Tilde{f(R, P)}=f(R)+g(P)=f_0 e^{n R}+ g_0 e^{m P}$, with $f_0, g_0, n, m$ constant parameters. In this case we shall introduce the following auxiliary variables by analyzing the Friedmann constraint equation:
\begin{equation}
    s=\frac{\rho_m}{3 H^2 (1+2 f'(R))},
\end{equation}

\begin{equation}
    x_1=\frac{f(R)}{3 H^2 (1+2 f'(R))},
\end{equation}

\begin{equation}
    x_2=\frac{R f(R)}{3 H^2 (1+2 f'(R))},
\end{equation}

\begin{equation}
    x_3=\frac{2 \dot{R} f''(R)}{H (1+2 f'(R))},
\end{equation}

\begin{equation}
    x_4=n H^2,
\end{equation}

\begin{equation}
   z=\frac{R}{H^2}.
\end{equation} 

\begin{equation}
    y_1=\frac{g(P)}{3 H^2 (1+2 f'(R))},
\end{equation}

\begin{equation}
    y_2=\frac{6 \beta H^3 \partial_t (g'(P))}{1+2 f'(R)}.
\end{equation}

\par 
If we take into account the exponential decomposition $\Tilde{f(R, P)}=f_0 e^{n R}+ g_0 e^{m P}$ the $x_2$ variable becomes a dependent component, while the remaining independent variables are the following: $[s, x_1, x_3, x_4, z, y_1, y_2]$. Then, the Friedmann constraint equation \eqref{eqfrcstr} has the following form:
\begin{equation}
   s=-\frac{3 \beta  m x_4^3 y_1 (z-6)}{n^3}+x_1 \left(1-x_4 z\right)+x_3+y_1+y_2+1,
\end{equation}
reducing the dimension of the corresponding dynamical system with one unit. If we take into account a pressure--less matter fluid $(w_m=0)$, then the second modified Friedmann relation can be written as:
\begin{multline}
    \frac{1}{3} \left(\frac{12 x_4}{n}-\frac{x_4 z}{n}\right)-\frac{3 x_4}{n}=-\frac{18 \beta  m^2 x_4^3 y_1 \ddot{P}}{n^3 \left(6 x_1 x_4-1\right)}
    \\-\frac{126 \beta  m x_4^4 y_1}{n^4 \left(6 x_1 x_4-1\right)}+\frac{15 \beta  m x_4^4 y_1 z}{n^4 \left(6 x_1 x_4-1\right)}
    \\-\frac{n^2 y_2^2}{18 \beta  m x_4^2 \left(6 x_1 x_4-1\right) y_1}
    -\frac{6 n x_1 x_4 \ddot{R}}{6 x_1 x_4-1}
    \\-\frac{2 x_4 y_2 z}{3 n \left(6 x_1 x_4-1\right)}+\frac{6 x_4 y_2}{n \left(6 x_1 x_4-1\right)}
    \\-\frac{3 x_4 y_1}{n \left(6 x_1 x_4-1\right)}+\frac{x_1 x_4^2 z}{n \left(6 x_1 x_4-1\right)}+\frac{6 x_1 x_4^2}{n \left(6 x_1 x_4-1\right)}
    \\-\frac{3 x_1 x_4}{n \left(6 x_1 x_4-1\right)}-\frac{2 x_3 x_4}{n \left(6 x_1 x_4-1\right)}-\frac{x_3^2}{6 n x_1 \left(6 x_1 x_4-1\right)}.
\end{multline}
\par 
In this case the final dynamical system obtained is described by the following differential equations:
\begin{equation}
\label{secprim}
    \frac{dx_1}{dN}=-\frac{x_1 z}{3}-x_3 x_1+4 x_1+\frac{x_3}{6 x_4},
\end{equation}

\begin{equation}
    \frac{dx_3}{dN}=6 n^2 x_1 \ddot{R}+x_3 \left(2-\frac{z}{6}\right)+x_3^2 \left(\frac{1}{6 x_1 x_4}-1\right),
\end{equation}

\begin{equation}
    \frac{dx_4}{dN}=\frac{1}{3} x_4 (z-12),
\end{equation}

\begin{equation}
    \frac{dz}{dN}=\frac{x_3}{6 x_1 x_4^2}-\frac{z^2}{3}+4 z,
\end{equation}

\begin{equation}
    \frac{dy_1}{dN}=\frac{n^3 y_2}{18 \beta  m x_4^3}-x_3 y_1-\frac{y_1 z}{3}+4 y_1,
\end{equation}

\begin{equation}
\label{secultim}
    \frac{dy_2}{dN}=\frac{18 \beta  m^2 x_4^2 y_1 \ddot{P}}{n^2}+\frac{n^3 y_2^2}{18 \beta  m x_4^3 y_1}-x_3 y_2+\frac{y_2 z}{2}-6 y_2.
\end{equation}
\par 
In order to close the dynamical system an additional relation between $\ddot{R}$ and $\ddot{P}$ is obtained, by differentiating the values specific to $R$ and $P$ components with respect to time. Considering the auxiliary variables, we obtain the following relation:
\onecolumngrid
\begin{multline}
    \ddot{P}=-\frac{10 n^2 y_2}{9 \beta  m^2 x_4^2 y_1}+\frac{n^2 y_2 z}{9 \beta  m^2 x_4^2 y_1}+\frac{1728 \beta  x_4^4}{n^4}-\frac{2 \beta  x_3 x_4^2}{n^4 x_1}-\frac{3 \beta  x_4^4 z^3}{n^4}+\frac{84 \beta  x_4^4 z^2}{n^4}-\frac{720 \beta  x_4^4 z}{n^4}+\frac{3 \beta  x_4^2 \ddot{R}}{n^2}.
\end{multline}
\twocolumngrid
\par 
At this point the second system of ordinary differential equations \eqref{secprim}--\eqref{secultim} is completely autonomous and can be analyzed using a dynamical system approach. In this case we have identified two critical points by analysing the right hand side of the \eqref{secprim}--\eqref{secultim} equations.
\par 
The first class of critical points is located at the following coordinates:
\onecolumngrid
\begin{equation}
    D=\Big[x_1= \frac{-18 \beta  m x_4^3 y_1+n^3 y_1+n^3}{n^3 \left(12 x_4-1\right)},x_3=0,z=12,y_2=0\Big].
\end{equation}
\twocolumngrid
\par 
The cosmological epoch associated to this class represents a de--Sitter era ($w_{eff}=-1$) where the geometrical dark energy component completely dominates in terms of density parameters ($s=0$). Moreover, we note that the auxiliary variables $y_1$ and $x_4$ have independent values. Furthermore, for this specific solution the values of the auxiliary variables $y_1$ and $x_4$ determine the location of the $x_1$ coordinate. From a physical point of view, this solution represents a geometrical de--Sitter epoch where the values of $n$ and $m$ parameters are affecting the location in the phase space structure and the dynamical properties. The eigenvalues of this solution have the following representations if we set some of the independent variables ($y_1=-1, x_4=1, \beta=-2$):
\onecolumngrid
\begin{equation}
\label{ecDpunct}
\Big[0,0,-3,4,-\frac{594 m^2 n^3\pm\sqrt{2} \sqrt{m n^6 \left(490050 m^3+84744 m^2 n^3+3775 m n^6+11 n^9\right)}+54 m n^6}{396 m^2 n^3+36 m n^6} \Big].
\end{equation}
\twocolumngrid
\par 
From a dynamical point of view the D solution is always saddle with one positive eigenvalue and one negative eigenvalue. We also note that the first two eigenvalues are equal to zero, describing a non--hyperbolic solution. Hence, the values of the $n$ and $m$ coefficients are affecting the behavior in the phase space for the spiral properties of the corresponding trajectory. A non--exclusive region where the trajectory in the phase space is spiral due to the complex behavior of the last two  eigenvalues in eq. \eqref{ecDpunct} is displayed in Fig.~\ref{figDa}.

\par 
Lastly, the second class of critical points is located at the following coordinates:
\onecolumngrid
\begin{equation}
    F=\Big[x_3=0,x_4=\frac{1}{12},z=12,y_1=-\frac{96 n^3}{96 n^3-\beta  m},y_2=0\Big].
\end{equation}
\twocolumngrid
\par 
The cosmological epoch associated to this class represents also a de--Sitter era ($w_{eff}=-1$) where the geometrical dark energy component completely dominates in terms of density parameters ($s=0$). For this solution we note that the auxiliary variables $x_1$ is independent. From a physical point of view, the F solution represents a geometrical de--Sitter epoch where the values of $n, m$ and $\beta$ parameters are affecting the location in the phase space structure through the value of the $y_1$ component. The eigenvalues of this solution are the following:
\begin{equation}
\Big[0,0,-3,4, \Xi_5,  \Xi_6 \Big].
\end{equation}
\par 
For this solution the specific form of the last two eigenvalues $\Xi_5,  \Xi_6$ is not displayed due to the complicated form of the obtained relations. From a dynamical point of view, the E solution has a similar behavior to the previous one, describing a non--hyperbolic epoch always saddle where the de--Sitter dynamics appears from the curvature and cubic extensions. As in the previous point, the values of the $n$ and $m$ coefficients are affecting only the spiral properties of the phase space. In the Fig.~\ref{figFa} we have displayed a possible region characterized by a spiral behavior due to the imaginary values of the last eigenvalues $\Xi_5$ and $\Xi_6$.

\section{Conclusions}
\label{sec:concluzii}
\par 
In this paper we have extended the Einstein-Hilbert action by adding a generic term $\Tilde{f}(R,P))$ which depends on the scalar curvature $R$ and the specific invariant $P$, based on different cubic contractions of the Riemann tensor. This action can be regarded as an attempt which corrects the $\Lambda CDM$ model by encoding particular geometrical invariants, leading to possible interesting dynamical effects. In our approach we have first considered that the generic term $\Tilde{f}(R,P)$ can be written as a direct sum between the geometrical constituents, taking into account the power law decomposition, i.e. $\Tilde{f}(R,P) \to f_0 R^n+g_0 P^m$. 
\par 
After we have obtained the specific modified Friedmann equations, the constraint and the acceleration equation, we have investigated the physical characteristics of the present cosmological model by considering an analytical approach based on the dynamical system analysis. Hence, we have introduced specific dimension--less variables, approximating the evolution of the cosmological model as an autonomous system of ordinary differential equations, applying the analytical methods associated to dynamical systems. In this regard, we have obtained the critical points specific to the present cosmological model, analyzing the location in the physical phase space and the viability of these solutions. For each critical point we have determined the associated eigenvalues, detecting the dynamical characteristics. In the case of the power law decomposition the present dynamical system  has two critical points. 
\par 
The first critical point represents a de--Sitter epoch, where the effective equation of state corresponds to a cosmological constant-like solution. Analyzing the eigenvalues corresponding to this critical point, we have noticed the this point cannot be stable or unstable, always representing a saddle solution, due to the existence of one eigenvalue with positive real part, and one eigenvalue with a negative real part. Moreover, due to the presence of one zero eigenvalue, we also note that this solution represents a non--hyperbolic equilibrium in the physical phase space.
\par 
The second critical point represents a cosmological era where the dynamical features depend on the specific values of the geometric couplings, the $n$ and $m$ parameters, encoding effects due to the curvature and the cubic gravity type parameterization. In this case the effective equation of state is sensitive to the values of the $n$ and $m$ parameters and can describe various epochs. Hence this solution can also be associated to a matter or radiation era for specific values of the $n$ and $m$ constant parameters. Moreover, it can describe also a quintessence or phantom-like epoch, explaining the super--acceleration scenario. Furthermore, it can also be associated to some cosmological solutions, stiff or super--stiff dynamics. For this solution the form of the eigenvalues depend on the specific values of the $n$ and $m$ parameters, describing a hyperbolic equilibrium. We have analyzed the second equilibrium point from a dynamical point of view, showing that in certain cases this solution can be stable, saddle, or unstable. Lastly, some specific solutions have been analyzed, determining the evolution and the viability of the analytical solutions.
\par 
In our study we have also considered an exponential decomposition $\Tilde{f}(R, P)\to f_0 e^{n R}+ g_0 e^{m P}$, with $f_0, g_0, n, m$ constant parameters in Sec.~\ref{sec:atreiab}. For this specific case we have analyzed the structure and properties of the corresponding phase space, revealing the cosmological solutions obtained. For the exponential model we have identified two cosmological de--Sitter epochs where the values of the $n$ and $m$ parameters are affecting the location in the phase space and the dynamical properties of the corresponding trajectories. 
\par 
The present paper offers a generalization to the $f(R)$ gravity theories, by embedding an invariant component denoted $g(P)$, based on specific contractions of the Riemann tensor in the third order. In this way the $f(R)$ gravity theory is extended, by including specific geometrical manifestations from the cubic component, offering a generalization to the basic Einstein-Hilbert action towards a more complete gravity theory. In the current manuscript the investigation is based on the usage of dynamical system analysis, an important tool in cosmology \cite{Odintsov:2017icc}. We mention here that in order to better discriminate between the effects due to the scalar curvature part and the cubic component, various reconstruction methods can in principle be applied \cite{Nojiri:2010wj} to the present model, obtaining possible constraints due to different dynamical behaviors. Another important aspect in the dynamical system analysis is related to the viability of the corresponding singularities at finite time. For the $f(R)$ gravity theory the properties of the resulting singularities have been investigated in Ref.~\cite{Odintsov:2017tbc}. In principle, such an analysis can be applied for specific models of $f(R,P)$, obtaining possible constraints from the physical properties of the corresponding models.
\par 
We note that the present cosmological model can be also extended in various applications, by analyzing different specific models and solutions to the gravitational field equations, by considering effects due to the curvature and the cubic term. One particular extension is related to the inclusion of the Gauss--Bonnet topological invariant, leading to a more generic theory of gravity. A different approach is related to the study of the inflationary era, an epoch where this theory can have visible effects in the early times. Furthermore, one can consider different specific models for the $\Tilde{f}(R,P)$ gravity by taking into account possible parameterizations, as an exponential decomposition or more complex models which can include a cosmological constant.  All of the previous mentioned questions and directions are open and left as possible future investigations.
\par 
Another important aspect in modified gravity theories is related to the study of various instabilities which can manifest. In the case of Einsteinian cubic gravity various studies \cite{Pookkillath:2020iqq, Jimenez:2020gbw} have indicated that some pathological instabilities might emerge, leaving the corresponding theory unhealthy from a physical point of view. The gravitational theory studied at the background level in the present paper represents a generalised attempt based on non--linear cubic and curvature extensions for the Einstein--Hilbert action. To this regard, the current model have to be considered only as an effective theory and needs to be analyzed from this point of view. An expanded discussion on this issue is presented in a recent publication \cite{Cano:2020oaa}.

\section{Acknowledgements}
  For this project we have considered different analytical computations in $Wolfram$ $Mathematica$ \cite{Mathematica} and $xAct$ \cite{xact}. The computational part of this work was performed utilizing the computer stations provided by CNFIS through the project CNFIS-FDI-2020-0355.

\bibliography{article}
\bibliographystyle{apsrev}

\end{document}